\documentstyle[aps,epsfig]{revtex}
\newcommand{\be}{\begin{equation}}
\newcommand{\ee}{\end{equation}}
\newcommand{\bea}{\begin{eqnarray}}
\newcommand{\eea}{\end{eqnarray}}

\def\simlt{\stackrel{<}{{}_\sim}}
\def\simgt{\stackrel{>}{{}_\sim}}
\begin{document}
\title{Amplification of hypercharge electromagnetic 
fields by a cosmological pseudoscalar}

\author{ Ram Brustein and David H. Oaknin }

\address{Department of Physics, Ben-Gurion University, 
Beer-Sheva 84105, Israel\\
email: ramyb,doaknin@bgumail.bgu.ac.il}

\maketitle{
\begin{abstract}

If, in addition to the standard model fields,
a new pseudoscalar field exists and couples to hypercharge
topological number density, it can exponentially amplify hyperelectric
and hypermagnetic fields in the symmetric phase of the electroweak
plasma, while coherently rolling or oscillating. We present the equations
describing the coupled system of a pseudoscalar field and hypercharge
electromagnetic fields in the electroweak plasma at temperatures above the 
electroweak phase transition,   discuss  approximations to the
equations, and their validity. We then solve the approximate equations
using assorted analytical and numerical methods, and determine 
the parameters for which hypercharge electromagnetic fields 
can be exponentially amplified.

\end{abstract} 
\centerline{Preprint Number: BGU-PH-98/13}

\section{INTRODUCTION}
The origin of the cosmological baryon asymmetry remains one of the most 
fundamental open questions in high energy physics, in spite of 
the effort and attention  it has attracted in the last three decades. 
In 1967 Sakharov noticed \cite{Sakharov} that three conditions are
essential for the creation of a net baryon number in a previously
symmetric universe: 1) baryon non-conservation; 2) C and CP violation; 3)
out of equilibrium dynamics . Since then many different hypothetical
cosmological scenarios in which the three conditions could be fulfilled
have been proposed as possible scenarios for baryogenesis. 

Among the different scenarios the electroweak (EW) scenario 
plays a leading role. It is particularly
appealing because it involves physics that can be
experimentally tested in the working colliders and those that will turn on
during the coming years.
Non-perturbative sphaleron processes, at thermal equilibrium at
temperatures above the EW phase transition,
erase any previously generated baryon excess along the $B-L=0$ direction. 
In addition, if some asymmetry is generated during the
transition it is erased immediately after the phase transition
completes by the same sphaleron processes, if the transition is not strong
enough to effectively suppress them \cite{shaposhnikov}. The strength of
the EW phase transition has been extensively studied in the Standard Model
(SM) and its popular extensions \cite{strength1,strength2}, including
leading quantum and thermal corrections to the finite temperature
effective potential. In the SM the phase transition seems to be second
order or even completely absent for those large values of the higgs mass
that have not been ruled out by LEP II experiment. In addition it became
clear that the mechanisms that were considered had difficulties to
generate enough asymmetry to explain the observed baryon to entropy ratio 
\cite{kaplan}. One of the most dramatic conclusions that emerged from
these studies was that either new physics beyond the SM able to change the
character of the phase transition and generate enough asymmetry was
relevant at the EW scale, or that new physics at much higher energies was
responsible for a generation of an asymmetry $B-L \neq 0$ that has 
survived until today.

It has been recently noticed \cite{gs} that hypermagnetic (HM) fields could
be significant players in the EW scenario for baryogenesis. Long range
uniform magnetic fields could strengthen the EW phase transition to the
point that it is strong enough even for the experimentally allowed values of
the SM higgs mass.  The reason is that only the projection of the
hyperfields along the massless photon can propagate inside the bubbles of
the broken phase, while their projection along the massive $Z$-boson
cannot propagate.  This well-known effect in conductor-superconductor
phase transition \cite{meissner} adds a pressure term to the symmetric
phase which can lower the transition temperature. A detailed study of this
effect in the SM phase transition has been attempted in several recent
papers \cite{enqvist}. The results are
not quite conclusive at the moment,  however, this effect could save a
baryon asymmetry with $B-L=0$ generated during the phase transition from
erasure by sphaleron transitions in the broken phase and, therefore, may
fix one of the two main SM dissabilities discussed above.

A subtle effect of hyperelectromagnetic (HEM) fields in the
EW scenario may also solve the other basic problem for EW
baryogenesis, the amount of asymmetry that can be generated.  Giovannini
and Shaposhnikov have shown \cite{gs} that the topological Chern-Simons
(CS) number stored in the HEM fields just before the
phase transition is converted into a fermionic asymmetry along the
direction $B-L=0$. 

We have shown \cite{bo1}, that an extra axion-like pseudoscalar field
coupled to the hypercharge topological number density can 
amplify HM fields in the unbroken phase of the EW plasma, while coherently
rolling or oscillating around the minimum of its potential. This mechanism
is capable of generating a net CS number that can survive until the
transition and  then be converted in a baryonic asymmetry in sufficient
amount to explain the observed baryon to entropy ratio.

Pseudoscalar fields with the proposed axion-like coupling appear
in several possible extensions of the Standard Model. They typically have
only perturbative derivative interactions and therefore vanishing potential at
high temperatures, and acquire a potential at lower temperatures through
non-perturbative interactions. Their potentials take the generic form
$V_0^4 V(\phi/f)$, where $V$ is a bounded periodic function 
characterized by two mass scales: a large $f$, which could be as high as 
the Planck scale, and a much smaller mass $m\sim V_0^2/f$, which
could be as low as a fraction of an eV, or as high as $10^{12}$ GeV. A
particularly interesting (pseudo)scalar mass range is the TeV range,
expected to appear if potential generation is associated with
supersymmetry breaking. Scalars with axion-like 
coupling to hypercharge electromagnetic fields were previously considered
in \cite{widrow,guendelman}. 

Amplification of ordinary electromagnetic
(EM) fields by a scalar field was discussed in \cite{widrow}.
In this paper we discuss in detail the dynamics of the pseudoscalar
field and how it drives HEM fields in the EW plasma at temperatures above
the phase transition. In section II. we present the basic equations and
several useful approximations to them. An analytical description of the
amplification of HEM fields by this mechanism in the different regions of
the parameter space is presented in section III. In section IV we compare
this analytical approach with numerical results. Our conclusions are
summarized in section V.

\section{ELECTRODYNAMICS DRIVEN BY AN AXION-LIKE PSEUDOSCALAR.}

In this section we discuss hypercharge electrodynamics in the unbroken
phase of the EW plasma coupled to a cosmological pseudoscalar
field. We assume that the universe is homogeneous and isotropic, and can
be described by a conformally flat metric $ds^2 = a^2(\eta) (d\eta^2 -
dx_1^2 - dx_2^2 - dx_3^2)$ where $a(\eta)$ is the scale factor of the
universe, and $\eta$ is conformal time related to cosmic time $t$ as
$a(\eta) d\eta= dt$.

In addition to the SM fields we consider an extra pseudoscalar field
$\phi$ with coupling to the $U(1)_Y$ hypercharge field strength
$\frac{\lambda}{4}\phi Y\widetilde{Y}$. The scalar field rolls or
oscillates coherently around the minimum of its potential, 
at temperatures $T \simgt 100~GeV$ just above the EW phase
transition, and drives the dynamics of the HEM fields. The coupling
constant $\lambda$ (which we will take as positive) has units of
mass${}^{-1}$ $\lambda\sim 1/M$. For
QCD axions, the two scales $M$ and $f$, which is the typical scale of
variation for the scalar fields, are similar $M\sim f$, but
in general, it is not always the case. As we are not considering here
any particular extension of the Standard Model we take $M$ to be a free
parameter, and in particular allow $f>M$. 

We also assume that the universe is radiation dominated already at some
early time $\eta=0$, at $T\simgt 100 GeV$, before  the scalar
field becomes relevant.  Once the scalar field starts
its coherent motion it can become the dominant source of energy
in the universe.  We will assume that after a short period of time
the scalar field decays, perhaps to (hyper)photons or other particles, so
the radiation dominated character of the universe is not significatively
altered.

The lagrangian density describing  HEM fields, coupled to the 
heavy pseudoscalar in the resistive approximation \cite{biskamp} of the
highly conducting EW plasma is 
\begin{equation}
\label{lagrangian}
{\cal L}= \sqrt{-g} \left(\frac{1}{2}
\nabla_{\mu}\phi\nabla^{\mu}\phi 
- V(\phi) - \frac{1}{4} Y_{\mu\nu} Y^{\mu\nu} - J_{\mu} Y^{\mu} -
\frac{\lambda}{4}\phi Y_{\mu\nu} \widetilde{Y}^{\mu\nu} \right) - \mu
\epsilon_{ijk} Y^{ij} Y^k,
\end{equation}
where $g = det(g_{\mu \nu})$ is the determinant of the metric tensor,
$\nabla_{\mu}$ denotes the covariant derivative, and  $J_{\mu}$ is the ohmic
current. 

The last term in (\ref{lagrangian}) takes into account the 
possibility that a fermionic chemical potential survives in the
unbroken phase of the EW plasma \cite{dhana}. A finite right electron 
density could survive if the higgs mediated electron chirality flip
interaction is sufficiently suppressed, due to the smallness of the
electron Yukawa coupling, so that it cannot reach thermal
equilibrium \cite{gs,cline}. The coefficient $\mu$ in ({\ref {lagrangian}}) 
is related to the right electron chemical potential $\mu_R$  by $\mu =
\frac{g'^2}{4 \pi^2} \mu_R$, where $g'$ is the hypercharge gauge coupling
constant. The latin indexes  ($i,j,k$) run over the spatial degrees of
freedom only. 

The equations of motion that follow from lagrangian (\ref{lagrangian})
have been partially derived by several authors \cite{gs,widrow,dettmann}.
The complete set of equations of motion includes an  equation for the
pseudoscalar,
\begin{equation}
\nabla^{\mu}\nabla_{\mu}\phi
+ \frac{dV(\phi)}{d\phi} = - \frac{\lambda}{4}\phi Y_{\mu\nu}
\widetilde{Y}^{\mu\nu}, 
\label{scalar}
\end{equation}
an equation  for HEM fields,
\begin{equation}
\nabla_{\mu} Y^{\mu\nu} = J^{\nu} + 4  \delta_{\mu}
\widetilde{Y}^{\mu\nu} - \lambda (\nabla_{\mu}\phi)
\widetilde{Y}^{\mu\nu},
\end{equation}
and the Bianchi identity,
\begin{equation}
\nabla_{\mu} \widetilde{Y}^{\mu\nu} = 0.
\end{equation}
We have introduced the four-vector $\delta_{\alpha}=(\mu,0,0,0)$
in order to keep the notation clear.

A plasma is a highly conducting ionized medium, so that
individual (hyper)electrical charges are exponentially
shielded $\sim e^{-r/{\it r}_D}$ over distances longer than the Debye
radius ${\it r}_D$, which in the EW plasma is of the order of
the inverse of the temperature of the plasma $T$, ${\it r}_D^{-1} \sim
(10-100) T$ \cite{baym}. Therefore, we
will  assume that the plasma is  electrically neutral over 
length scales larger than ${\it r}_D$.
Since we are interested in the coherent motion of the time-dependent
scalar field we will assume, in addition, that the spatial derivatives of
$\phi$ are negligible compared to the other terms in the equations.

Maxwell's equations for HEM fields then become
\begin{eqnarray}
&(i)&\ \nabla \cdot {\vec B} = 0 
\nonumber \\
&(ii)&\ \frac{\partial {\vec B}}{\partial \eta} = -\nabla \times {\vec E}
\nonumber \\
&(iii)&\ \nabla \cdot{\vec E} = 0 
\nonumber \\
&(iv)&\ \frac{\partial {\vec E}}{\partial \eta} =  \nabla \times {\vec B}
+ \frac{g'^2}{\pi^2} \mu_R a(\eta) {\vec B} - \lambda \frac{d \phi}{d
\eta} {\vec B} - {\vec J},
\label{maxwell}
\end{eqnarray}
where $\nabla$ now represents the usual three-space gradient (for
comoving coordinates). The current ${\vec J}$ is given by Ohm's law
in a plasma whose bulk motion is described by a non-relativistic velocity
field ${\vec v}$, 
\begin{equation}
{\vec J} = \sigma ({\vec E} + {\vec v} \times {\vec B}).
\label{ohm}
\end{equation}

In  equations (\ref{maxwell},\ref{ohm}) we have introduced 
rescaled electric and magnetic fields 
$\vec E=a^2(\eta) \vec{\cal E}$, $\vec B=a^2(\eta) \vec{\cal B}$,  
and physical conductivity $\sigma=a(\eta)\sigma_c$. The fields $\vec{\cal
E}$, $\vec{\cal B}$ are the flat space HEM fields defined through  
\begin{eqnarray}
Y^{\mu\nu} = a^{-2} \left( 
\begin{array} {cccc}
         0        & {\cal E}_{\it x} & {\cal E}_{\it y} & {\cal E}_{\it z}
\\
-{\cal E}_{\it x} &         0        & {\cal B}_{\it z} &-{\cal B}_{\it y}
\\
-{\cal E}_{\it y} &-{\cal B}_{\it z} &         0        & {\cal B}_{\it x}
\\
-{\cal E}_{\it z} & {\cal B}_{\it y} &-{\cal B}_{\it x} &        0 
\end {array}
\right).
\end{eqnarray}

Equation ({\ref {scalar}}) for the scalar $\phi$ becomes, after expanding
the covariant derivative, 
\begin{equation} 
\label{axion} 
\frac{d^2 \phi}{d
\eta^2} + 2 a H \frac{d \phi}{d \eta} + a^2 \frac{d{\em V}(\phi)}{d\phi} =
\lambda a^2 {\vec E} \cdot{\vec B}.
\end{equation}
The Hubble parameter $H=\frac{1}{a^2} \frac{da}{d\eta}$
is related to the temperature $T$,
$
H=\frac{T^2}{M_0}
$, where $M_0 =  \frac{M_{Pl}}{1.66 \sqrt{g_{*}}}$, and $g_{*}$ is the
number of effective degrees of freedom in the thermal bath of the plasma.
At the EW scale $g_{*}=106.75$ for the SM degrees of freedom in the
unbroken phase. For us, although the difference is not very relevant
in our analysis, $g_{*}=107.75$ if we consider the extra 
bosonic degree of freedom.

In the absence of the extra pseudoscalar field, Maxwell's equations in 
a plasma can be solved in the magnetohydrodynamics approach (MHD) of
infinite conductivity $\sigma \rightarrow \infty$ \cite{biskamp},
\begin{eqnarray}
\frac{\partial {\vec E}}{\partial \eta} \approx 0, \\
{\vec E} \approx  -{\vec v} \times {\vec B}.
\label{orthogonality}
\end{eqnarray}
In this approximation the electric ${\vec E}$ and magnetic fields ${\vec 
B}$ are perpendicular to each other, so that ${\vec E}\cdot{\vec B}=0$. 
We can therefore solve eq.({\ref {axion}}) with a vanishing r.h.s., as a
first approximation, and substitute the resulting $\phi(\eta)$ into
eq.({\ref{maxwell}{\em .iv}) as a background. At the end of the
section we give an estimate of the error coming from
neglecting the HEM backreaction in the scalar field equation. 

The explicitly time dependent background spoils, however, the
validity of MHD approach to Maxwell's equations. We therefore need a
generalized {\it ansatz} to describe the solutions to these equations. 
Our guess is that the HE field is proportional to the time derivative of the
HM field,
\begin{equation}
{\vec{E}} = \frac{\alpha}{\sigma} 
\frac{ \partial{\vec{B}} } { \partial\eta } 
+ {\vec{\nabla}}\theta,
\label{ansatz}
\end{equation} 
where $\alpha$ is a dimensionless coefficient to be determined.  We add
the term ${\vec \nabla}\theta$ for generality and consistency.  In the
choice of this {\it ansatz} we have been guided by two requirements:

1) when the scalar field decouples, the new
{\em ansatz} should be the leading correction in
the small parameter $\frac{1}{\sigma}$ to the MHD solution; 

2) the HE field ${\vec{E}}$ depends  linearly on the 
HM field  ${\vec{B}}$. 

Equations (\ref {maxwell}{\em.i}) and (\ref {maxwell}}{\em .iii}) can be
understood as constraints on the 
initial conditions, which are conserved in time when
${\vec E}$ and ${\vec B}$ evolve according to the other two  equations. 
Inserting ({\ref {ansatz}}) into  ({\ref {maxwell}}.{\em ii}) 
we obtain 
\begin{eqnarray}
\nabla \times {\vec B} &=& - \frac{\sigma}{\alpha} {\vec B}
\label{rotorB}
\\
\nabla \times {\vec E} &=& - \frac{\sigma}{\alpha} {\vec E} + 
\frac{\sigma}{\alpha} {\vec \nabla}\theta. 
\label{rotorE}
\end{eqnarray}
The last relation is consistent with 
({\ref {maxwell}}.{\em iii}), only if we require, in addition,  
$\nabla^2 \theta=0$. 
Now we have the same constraint on the spatial dependence of both 
${\vec B}$ and  ${\vec E}$ fields. This constraint can be 
most easily solved by assuming a solution with factorized
time dependence.  
The spatial modes that satisfy  eq.({\ref {rotorB}}) can be labeled by 
their wave number ${\vec k}$ and a sign $\pm$ for the two possible
helicities,
\begin{equation}
{\vec B}_{\vec k} = e^{-i{\vec k}{\vec x}} {{\vec b}_{\vec k}}{}^{\pm}\
\beta^\pm_{\vec k}(\eta), 
\label{modesB}
\end{equation}
with 
\begin{equation}
{{\vec b}_{\vec k}}{}^{\pm}= b_k^\pm  ({\hat e}_1 \pm i{\hat e}_2).
\label{polarization}
\end{equation}
Here ${\hat e}_1$, ${\hat e}_2$ are unit vectors in the plane
perpendicular to ${\vec k}$ such that $({\hat e}_1,{\hat e}_2,{\hat k})$
is a right-handed system. When we compute the curl of these spatial modes
we find that the proportionality constant $\alpha$ in our guess
({\ref {ansatz}}), for each given mode, should be  
\begin{equation} 
\alpha = \pm \frac{\sigma}{k}.
\label{proportionality}
\end{equation}

Once we have identified the magnetic modes  we
can, in a straightforward way, obtain the corresponding electric modes 
from eq. ({\ref {ansatz}}),
\begin{equation}
{\vec{E}}_{\vec{k}} = e^{-i{\vec k}{\vec x}} {{\vec{e}}_{\vec{k}}}{}^{\pm}\
\epsilon^\pm_{\vec k}(\eta) + {\vec \nabla}\theta,
\label{modesE}
\end{equation}
with
\begin{equation}
k {{\vec e}_{\vec k}}{}^{\pm} \epsilon^\pm_{\vec k}(\eta)= 
\pm {{\vec b}_{\vec k}}{}^{\pm}  
\frac{\partial \beta_{\vec k}}{\partial\eta}^\pm.
\label{modesEB}
\end{equation}

When we substitute expressions (\ref{modesB},\ref{modesE}) into 
eq.({\ref{maxwell}}.{\em iv}) 
we obtain a second order equation for the time dependent factors 
$\beta^\pm_{\vec {k}}(\eta)$, 
\begin{equation}
\label{evolution}
\frac{\partial^2  \beta_k^\pm }{\partial {\eta}^2}+
\sigma
 {\frac{\partial \beta_k}{\partial \eta}}^\pm + 
\left( k^2 \pm \lambda \frac{d \phi}{d \eta} k \pm \sigma \gamma k
\mp \frac{g'^2}{\pi^2} \mu_R k a(\eta) \right)
\beta_k^\pm(\eta) = 0.
\end{equation}
Equation (\ref{evolution}) is the main result of this section. 
The rest of the paper is devoted to finding useful approximations
to it, analyzing their solutions, and to numerical studies of its solutions.

The term $\pm \sigma \gamma k \beta_k^\pm(\eta)$ is the so-called
dynamo-term. Such a term is not unusual in plasma physics. It has been
claimed \cite{zeldovich} that such a term  could give rise to the long-range
homogeneous magnetic fields  in astrophysical
systems. A similar effect is provided by the $\mu$-term \cite{gs}. The
coefficient $\gamma$ is defined in the approximation 
\begin{equation}
\nabla \times ({\vec v} \times {\vec B}) \approx \gamma (\nabla \times
{\vec B}),
\label{dynamo}
\end{equation}
according to the procedure outlined in \cite{zeldovich}. If the bulk
motion of the plasma is random
and has zero mean velocity $<{\vec v}> = 0$, then it is possible to
average over the possible velocity fields,  assuming that the 
correlation  length of the magnetic field is much larger than the
correlation length of the velocity field. The procedure is equivalent to
averaging over scales and times exceeding the characteristic correlation
scale and time $\tau_0$ of the velocity field. On the other hand, if the
characteristic correlation scale and time of the velocity field are much
larger than those of the magnetic field we can assume ${\vec v}=0$. The
coefficient $\gamma$ is given by \begin{equation}
\gamma = -\frac{\tau_0}{3}
\langle{\vec v} \cdot {\vec \nabla} \times {\vec v}\rangle,
\end{equation}
and so, vanishes in the absence of vorticity in the plasma bulk motion.
{}From eq. ({\ref {dynamo}}) it is possible to conclude that
\begin{equation}
{\vec v} \times {\vec B} \approx \gamma {\vec B} + {\vec \nabla}\omega
\end{equation}
for a certain given scalar function $\omega$. 

The consistency of the 
analysis that we have presented imposes
\begin{equation}
\frac{\partial {\vec \nabla}\theta}{\partial \eta} + \sigma 
{\vec \nabla}\theta + \sigma {\vec \nabla}\omega = 0.
\label{escalar}
\end{equation}
This constraint can be obtained if we subtract ({\ref {maxwell}}.{\em ii})
from ({\ref {maxwell}}.{\em iv}) and then insert  our particular solution.
In the limit of infinite conductivity $\sigma \rightarrow \infty$ and in
the absence of vorticity ($\gamma=0$) we have
\begin{equation}
{\vec \nabla}\theta = -{\vec \nabla}\omega = -({\vec v} \times {\vec B}) +
\gamma {\vec B} = -{\vec v} \times {\vec B}.
\label{novorticity}
\end{equation}
In the particular case when the extra pseudoscalar is not present our {\it
ansatz} describes exact modes for electrodynamics in a conducting
medium. 

When the extra pseudoscalar is present, an exact analysis should
take into account the backreaction of the electromagnetic modes on the
scalar equation (\ref {axion}). We can estimate the error in
dropping this backreaction. 

The concrete scalar potential that we use for the analysis is
\be
V(\phi)=V_0^4 V(\phi/f)=m^2 f^2 V(\phi/f) \approx \frac{1}{2} m^2 \phi^2,
\ee
for $\phi \simlt f$. For values of $m$ in the TeV range and 
$T\simgt 100 GeV$, the mass term in ({\ref {axion}}) $a^2
\frac{d{\em V}(\phi)}{d\phi} =  a^2 m^2 \phi$ is dominant over the cosmic
friction term $ 2 a H \frac{d \phi}{d \eta}$. If we drop the
electromagnetic backreaction, the scalar field equation is that of
an harmonic  oscillator whose period$\sim 2\pi/m$,
is much shorter than the characteristic expansion time of the universe at
that time, $\frac{1}{H} \sim M_{Planck}/T^{2}$. Consequently, 
we can fix $a(\eta) = 1$ for the period of time that scalar oscillations
last. The general solution for the scalar field in this case is
\begin{equation}                                                               
\phi(\eta) =A cos(m\eta + \rho_0) \sim f cos(m\eta + \rho_0),  
\label{background}
\ee                                           
where $\rho_0$ is to be fixed as an initial condition. 

The backreaction term is negligible if 
\be
|\lambda {\vec E} \cdot{\vec B}| < |\frac{d^2 \phi}{d\eta^2}|.
\label{neglect}
\ee 

In a previous paper \cite{bo1} we have found, under very general
assumptions and using our particular {\it ansatz} to describe the
electrodynamics driven by an oscillatory scalar background, that
after averaging over magnetic and scalar fields initial conditions, 
$\langle{\vec E} \cdot{\vec B}\rangle$ can be expressed
as follows
\be
\langle{\vec E} \cdot{\vec B}\rangle = \frac{1}{2k_{\it
max}}\frac{\partial}{\partial
\eta} \langle{\vec B_{k_{\it max}}}^{+}{}^2-{\vec B_{k_{\it
max}}}^{-}{}^2 \rangle
\ee
where ${\vec B_{k_{\it max}}^{\pm}}$ is a particular mode that is
maximally amplified. Below we will show that $k_{max}$ is approximately
\be
k_{\it max} \approx \frac{1}{2} \lambda f m
\label{kmax}
\ee

If we integrate over a scalar oscillation we obtain average values over
time. Backreaction is negligible if 
\be
\frac{\lambda}{2k_{\it max}} |\langle{\vec B_{k_{\it max}}}^{+}{}^2-{\vec
B_{k_{\it max}}}^{-}{}^2\rangle| \ll |\langle\frac{d \phi}{d
\eta}\rangle| \sim f m.
\ee
Multiplying both sides of the equation by $f m$ and dividing
 by 
 $|\langle{\vec B_{k_{\it max}}}^{+}{}^2+
 {\vec B_{k_{\it max}}}^{-}{}^2\rangle|$, we obtain
\be
\frac{\lambda f}{2k_{\it max}/m} \frac{|\langle{\vec B_{k_{\it 
max}}}^{+}{}^2-{\vec B_{k_{\it max}}}^{-}{}^2\rangle|}{|\langle{\vec
B_{k_{\it max}}}^{+}{}^2+{\vec B_{k_{\it max}}}^{-}{}^2\rangle|} \ll
\frac{f^2 m^2}{|\langle{\vec B_{k_{\it max}}}^{+}{}^2+{\vec B_{k_{\it
max}}}^{-}{}^2\rangle|}
\ee
The right hand side of the inequality is less than unity  if the
energy density in the amplified magnetic modes exceeds that stored in the
scalar oscillations. So the condition can be estimated as 
\be
\frac{|\langle{\vec B_{k_{\it max}}}^{+}{}^2-{\vec B_{k_{\it
max}}}^{-}{}^2\rangle|}{|\langle{\vec
B_{k_{\it max}}}^{+}{}^2+{\vec B_{k_{\it max}}}^{-}{}^2\rangle|} \ll
\frac{2k_{\it max}/m}{\lambda f} \approx 1,
\ee
according to ({\ref {kmax}}). 

The asymmetry parameter 
\be
\label{asymmetry}
\gamma_B = \frac{\langle{\vec B_{k_{\it max}}}^{+}{}^2-{\vec B_{k_{\it
max}}}^{-}{}^2\rangle}{\langle{\vec B_{k_{\it 
max}}}^{+}{}^2+{\vec B_{k_{\it max}}}^{-}{}^2\rangle},
\ee 
can vary from $0$ - no asymmetry - to $\pm 1$ - total 
asymmetry. As we will show below (see also  
\cite{bo1}), in the general case of an oscillating scalar field both 
modes ($\pm$) are roughly amplified by the same exponential factor, so
that $\gamma_B$ remains approximately equal to its initial value. There we
also showed that this asymmetry parameter may be related to the amount
of baryonic asymmetry generated at the EW phase transition. We
concluded that only when this parameter is significatively small the
generated asymmetry is consistent with the experimental $n_B/s \sim
10^{-10}$. This justifies why we may neglect the electromagnetic
backreaction in the scalar equation.

\section{MAGNETIC FIELD AMPLIFICATION: ANALYTIC DESCRIPTION}

Before trying a 
detailed description of the general solution of eq.(\ref{evolution}), it
is useful to consider the simpler case of a constant $\frac{d \phi}{d
\eta}$. The solutions are then simply linear superposition of two
exponentials
\be
\beta^\pm(\eta) = \beta_1^\pm\ \hbox{\it\large e}^{\hbox{$\omega_1^\pm\eta$}}
+ \beta_2^\pm\ \hbox{\it\large e}^{\hbox{$\omega_2^\pm\eta$}},
\label{superposition}
\ee
where $\omega_{1,2}^{\pm}$ are the two roots of the quadratic equation,
\be
\label{roots}
\omega^2 + \sigma \omega + 
\left(k^2 \pm \lambda \frac{d \phi}{d \eta} k \pm \sigma \gamma k   
\mp \frac{g'^2}{\pi^2} \mu_R k a(\eta)\right) = 0,
\ee
which are 
\be
\omega_{1,2}^\pm =
\frac{1}{2} \left[
-\sigma \pm \sqrt{\sigma^2 - 4\left(k^2\pm \lambda \frac{d \phi}{d \eta}
k \pm \sigma \gamma k \mp \frac{g'^2}{\pi^2} \mu_R k a(\eta)\right) } 
\right].
\label{omega}
\ee

Unless a particular choice for the initial conditions is made, 
the solution is largely dominated by the exponential 
which corresponds to the larger of the two eigenvalues. This is an
exponentially  growing
function if $\omega_1$ is positive. The linear terms in the wave number
$k$ can exponentially amplify one of the two helicity modes for a limited
wave number values. For larger wave numbers the quadratic
term $k^2$ dominates over the linear terms and the magnetic modes are
oscillating and/or exponentially damped, depending on the overall sign of 
the expression under the square root. 

The right electron chemical
potential $\mu_R$ can be computed \cite{gs}
through the expression
\be
\mu_R = \frac{2}{45} \pi^2 g_{*} [\frac{783}{88}\delta_R - \frac{201}{88}
\delta_1 + \frac{15}{22}(\delta_2 + \delta_3) ] T
\label{chemical},
\ee
where $\delta_R=n_R/s$ ($s$ is the entropy density) is the right
electron
asymmetry. For generality we have included the possibility of finite
asymmetries $\delta_{i}$ in the three  conserved
charged $L_{i} - \frac{B}{3}$. The index $i=1,2,3$ runs over the three
fermion families. If we assume that these asymmetries can be of the order
of the observed baryonic asymmetry $n_B/s \sim 10^{-10}$ \cite{cline} we
get that the linear term in $k$, $\frac{g'^2}{\pi^2} \mu_R a(\eta)$, in
eq. ({\ref {omega}}) is of the order $10^{-10} T$, where $T$
 is the temperature of
the plasma. In the model that we are exploring the scalar velocity term
$\lambda \frac{d \phi}{d \eta} \sim \lambda f m > m$
is much larger than the chemical potential contribution. We will then fix
for simplicity $\mu_R=0$ in the following. 

We will assume that the velocity field ${\vec v}$ is parity invariant,
so that no vorticity at the length scales $k^{-1}$ we are considering is
present. We then fix $\gamma=0$ in addition to $\mu=0$.
Therefore, amplification can occur for one of the two
helicity modes if  
\be
\lambda\left| \frac{d \phi}{d \eta} \right|> k. 
\label{amplif}
\ee
To obtain significant amplification, coherent scalar
field velocities $\frac{d \phi}{d \eta}$ over a duration are
necessary, larger velocities leading to larger amplification.

It is interesting to remark at this point that, in absence of the
scalar field, and in the limit $\sigma \rightarrow \infty$, the
magnetic modes (both helicities)  diffuse,
\be
{\vec B}_k(\eta) \sim e^{-\frac{k^2}{\sigma}\eta}, 
\label{diffusion}
\ee
as MHD predicts \cite{gs}. Therefore, (see eq. (\ref {ansatz}) 
and (\ref{novorticity}))
\be
{\vec E}(\eta) = -{\vec v} \times {\vec B} \pm \frac{1}{k}
\frac{\partial {\vec B}}{\partial \eta} \approx -{\vec v} \times {\vec B}
\mp \frac{k}{\sigma} {\vec B},
\ee
that confirms, as we had anticipated, that in the appropriate limit our
exact modes describe the leading correction in $1/\sigma$ to 
the MHD approach ({\ref {orthogonality}}). 

In the general case of an oscillating scalar field ({\ref {background}}),
the field's velocity changes sign periodically and both modes can be
amplified, each during a different part of the cycle. Let  us first redefine 
eq. ({\ref {evolution}}) in terms of the dimensionless parameter $u = m
\eta$,  
\begin{equation}                                                               
\label{evolution2}                                      
\frac{\partial^2  \beta_k }{\partial u^2}^\pm+ \frac{\sigma}{m}          
\frac{\partial \beta_k}{\partial u}^\pm +                                  
\left( \left(\frac{k}{m}\right)^2 \pm \Lambda sin(u + \rho_0)
\left(\frac{k}{m}\right) \right)     
\beta_k^\pm(u) = 0,                                                      
\end{equation}          
where we have introduced the dimensionless parameter 
$\Lambda=\lambda A$. We can expect, on the basis 
of the example described at the beginning
of the section, that each of the two modes will grow
 exponentially  during parts 
of the scalar oscillation when  
\begin{equation}
\left(\frac{k}{m}\right) \pm \Lambda sin(u + \rho_0) < 0,
\label{fraction}
\end{equation} 
respectively, and  oscillate or
be damped during the other part of the
cycle. Equation ({\ref {fraction}}) imposes an upper bound on the wave
number spectrum that can get amplified, since $-1<sin(u)<+1$, we need that 
\begin{equation}
k < \Lambda m.
\label{upperbound}   
\end{equation} 

Net amplification results when growing overcomes damping during a cycle,
total amplification is then exponential in the number of cycles. 
In order to estimate in which cases this will happen, let us assume that $k$ is
small enough, $ k/m  < \Lambda$. Then, 
according to ({\ref {fraction}}), each mode gets amplified during
half a cycle and damped during the other half. If, in addition,
$4\left|\pm \Lambda sin(u + \rho) \left(\frac{k}{m}\right) \right| \ll
\left(\frac{\sigma}{m}\right)^2$, the relevant eigenvalue
$\omega_1$ can be approximated by a linear expression
\be
\omega_{1}^\pm \approx - \left(\pm \Lambda sin(u + \rho) \frac{k}{\sigma} 
\right),
\label{omega1}
\ee
so that the growth during half a
cycle is canceled by the damping when the sine changes its sign, and
therefore  no net
amplification results. But, if the two terms in the last inequality are at
least comparable, the linear approximation is not valid; a quadratic
correction, that does not change its sign with the sine function, is
relevant and net amplification results. 
This gives us a lower bound on the amplified wave number spectrum,
\begin{equation}
\label{lowerbound}
\Lambda \left(\frac{k}{m}\right)  \simgt
\left(\frac{\sigma}{m}\right)^2.
\end{equation}

We conclude that amplification occurs for a limited range in the wave
number spectrum. Maximum amplification occurs for that value of $k$
that maximaizes the expression $\left(\frac{k}{m}\right)^2 \pm \Lambda
\left(\frac{k}{m}\right) sin(u)$, namely,
\begin{equation}
k_{\it max} \sim \frac{\Lambda}{2} m.
\label{maximum}
\end{equation} 
To be more precise, this estimate is an upper bound on $k_{\it max}$. 
We have not taken into account that the larger $k$ is, the 
smaller the duration in  which  amplification occurs (see condition ({\ref 
{fraction}})). Nevertheless, numerical analysis confirms that ({\ref
{maximum}}) is quite accurate. (For example, see 
Fig. \ref{fig:amppcwn}).

The conductivity of the plasma $\sigma$, as well
as the comoving wave number $k$ are both proportional to the temperature
of the plasma $T$, $\sigma=\sigma_0 T \sim 10 T$ \cite{baym} and $k=k_0
T$. The character of
the solutions of eq.(\ref{evolution2}), for a given comoving wave number, 
depends on the dimensionless parameter $T/m$. In Fig. \ref{fig:amppct} we
show that for a
given $k_0$ there is a limited range of temperatures where the oscillating  
scalar field drives net magnetic amplification, as can be seen from eqs.
({\ref {upperbound}}) and ({\ref {lowerbound}}).

The situation is somewhat different when $m \ll T$. In this case
eq. (\ref{evolution2}) can be approximated by a first order equation 
\be
\frac{\sigma}{m}\frac{\partial \beta}{\partial u}^\pm + 
\left( \left(\frac{k}{m}\right)^2 \pm \lambda \frac{d \phi}{d u}
\left(\frac{k}{m}\right) \right) \beta^\pm(u) = 0,
\ee
which can be solved exactly, 
\be
\beta^{\pm}(\eta)= \beta^{\pm}
_0
\ \hbox{\it\large e}^{\hbox{$-\frac{k^2}{\sigma}\eta \mp \lambda
(\phi(\eta)-\phi(0))\ k/\sigma$}}.
\label{rollsol}
\ee
In this extreme limit, the
friction term in the scalar equation ({\ref {axion}}) is
comparable to the mass term. It means that it would take a time
interval of the order of the characteristic cosmic time expansion for
the scalar velocity to change  its value significatively,  so for our
purposes, we may simply consider
$d\phi/d\eta \approx constant$.  The scalar
field does not oscillate during the relevant time scale, instead we say
the scalar field rolls. When the scalar field rolls only one of the two
helicity modes gets amplified. The helicity mode that is amplified,
is determined by the sign of $\phi(\eta)-\phi(0)$. 
The amplification factor
${{\cal A}^\pm}(k,\eta)=\beta^\pm(\eta)/\beta^\pm(0)$ due to this
mechanism is maximal for the wave number $k$, that at a given time
$\eta$, maximizes the exponent in eq. ({\ref {rollsol}}): 
\be
k_{max}\eta=\frac{1}{2} \lambda \Delta \phi,
\label{rollwave}
\ee
\be 
{{\cal A}^\pm}(k_{max},\eta)= e^{ \frac{1}{4} 
(\lambda \Delta \phi)^2 \frac{1}{\eta\sigma}}.
\label{rollampl}
\ee
Here $\Delta \phi(\eta)=|\phi(\eta)-\phi(0)|$. Looking at $\eta\sim
\eta_{EW}$ we obtain $\frac{1}{\eta_{EW}\sigma}\sim
10^{-16}$, and therefore to obtain amplification $\lambda \Delta
\phi\simgt 10^8$. So we have
\be
\frac{k_{max,EW}}{T_{EW}} \approx 10^8 \cdot 10^{-16} \cdot
\frac{\sigma_{EW}}{T_{EW}} 
\approx 10^{-7}.
\ee

 A value
of $\lambda \Delta \phi\simgt 10^8$ is not unnatural,
for example, such a value is obtained if the typical scale for scalar field
motion is the Planck scale, as happens in many models of supergravity, and
$\lambda\simlt 1/10^{10} GeV$.

In the rolling case the discussion about the role of the electromagnetic 
backreaction on the scalar equation ({\ref {axion}}) is somewhat different
from the  oscillating case discussed at the end of the previous section.
As we have just shown only one of the two helicity modes is
amplified when the scalar field rolls, while the other is damped. The
backreaction term $\lambda a^2 {\vec E}\cdot{\vec B}$ can be expressed in
terms of the solution ({\ref {rollampl}}) 

\be
|\lambda a^2 {\vec E}\cdot{\vec B}| \approx \frac{\lambda}{2k_{\it max}}
a^2
|\frac{\partial}{\partial \eta}(\beta_{k_{\it max}}^2(\eta))| =
\frac{\lambda}{2k_{\it max}} a^2 |\frac{\partial}{\partial \eta}(\beta_0^2
e^{\frac{1}{2} (\lambda \Delta \phi)^2 \frac{1}{\eta\sigma}})|, 
\ee
where $k_{\it max}$ is the mode specified in ({\ref {rollwave}}).

This term is to be compared to any of the dominant terms in the scalar
equation ({\ref {axion}}). In this case we choose for convenience the
friction term $2 a H \frac{\partial \phi}{\partial \eta}$. The
backreaction is negligible while
\be
\frac{\lambda}{2k_{\it max}} a^2 \frac{\partial}{\partial \eta}(\beta_0^2 
e^{\frac{1}{2}(\lambda \Delta \phi)^2 \frac{1}{\eta\sigma}}) < 2 a H
\frac{\partial \phi}{\partial \eta}.
\ee
If we integrate over time then the condition can be written 
\be
\frac{\lambda}{4k_{\it max}} \frac{\beta_0^2 e^{ \frac{1}{2}(\lambda
\Delta   
\phi)^2 \frac{1}{\eta\sigma}}}{H\Delta \phi} < 1.
\label{bnd}
\ee

When  bound (\ref{bnd}) is saturated the electromagnetic backreaction
in the scalar field equation becomes relevant and changes
 the free motion of the scalar field. 

A complete discussion of the end of the coherent oscillations or rolling
by the scalar field is beyond the scope of this paper.
However, we would like to make two comments in this regard,
\begin{itemize}
\item
we do know that once the oscillations or rolling stop, the fields
are no longer amplified and obey a diffusion equation ({\ref
{diffusion}}). Modes
with wave number below the diffusion limit $k <k_\sigma T\sim 10^{-8} T$, where
$\frac{k_{\sigma}^2}{\sigma}\frac{1}{\eta_{EW}}=1$, remain almost
constant until the EW transition, their amplitude goes down as $T^2$, and
energy density as $T^4$, maintaining a constant ratio with the environment
radiation. Modes with $k/T>k_\sigma$ decay
quickly, washing out the results of amplification. We have seen that the
range of amplified momenta for oscillating fields is not
too different than $T$, therefore scalar field  oscillations have to occur
just before, or during the EW transition, for the mechanism we are
discussing in this paper to be relevant for EW baryogenesis. In that case,
the amplified fields do not have enough time to be damped by diffusion. If
the field is rolling, momenta $k\ll T$ can be amplified and then frozen
in the plasma until the phase transition, and therefore the rolling can
end sometime before the transition. 
\item
Some clues about how oscillations or rolling may eventually 
end can be obtained from the estimated HEM backreaction term 
in the scalar field equation. For an oscillating field, 
we have seen that the backreaction term remains negligible 
throughout the evolution. In that case, another type of effect or 
interactions have to be considered as leading eventually to the end of coherent
motion. If the field is rolling we have seen that the backreaction term can
become significant, and therefore it may well be that in this case
the coherent decay of the scalar field into HEM fields is the 
cause for the end of coherent motion.
\end{itemize}

\section{NUMERICAL ANALYSIS.}

We have studied the numerical solutions of eq.(\ref{evolution2}) 
in different regions of the parameter space. We find results that are in
very good agreement with the previous qualitative discussion:
amplification occurs for a limited range of
Fourier modes, peaked around  $k/m \sim \frac{1}{2} \Lambda$, (see eq.
(\ref {maximum}) and Fig. \ref{fig:amppcwn}). The modes of the EM fields
in
the
spectrum range that is amplified are oscillating with (sometimes
complicated) time dependence and an exponentially growing amplitude, 
\be
\beta_k^{\pm}(\eta)={\it\large e}^{\Gamma_k^{\pm}\frac{m\eta}{2\pi}}
{\it P}(\eta).
\label{perioampl}
\ee
The coefficient $\Gamma$ gives the amplification rate per oscillation of
the corresponding mode. The function ${\it P}(\eta)$ is periodic over a
period $\Delta\eta=2\pi/m$. 

In Fig. \ref{fig:amp}  we show 
the time dependence of three representative magnetic modes for specific 
values of parameters: (1) $k$ outside the amplification region; 
(2) low $k$ values; (3) $k$ values around maximum amplification.

\begin{figure}
\begin{center}
\psfig{figure=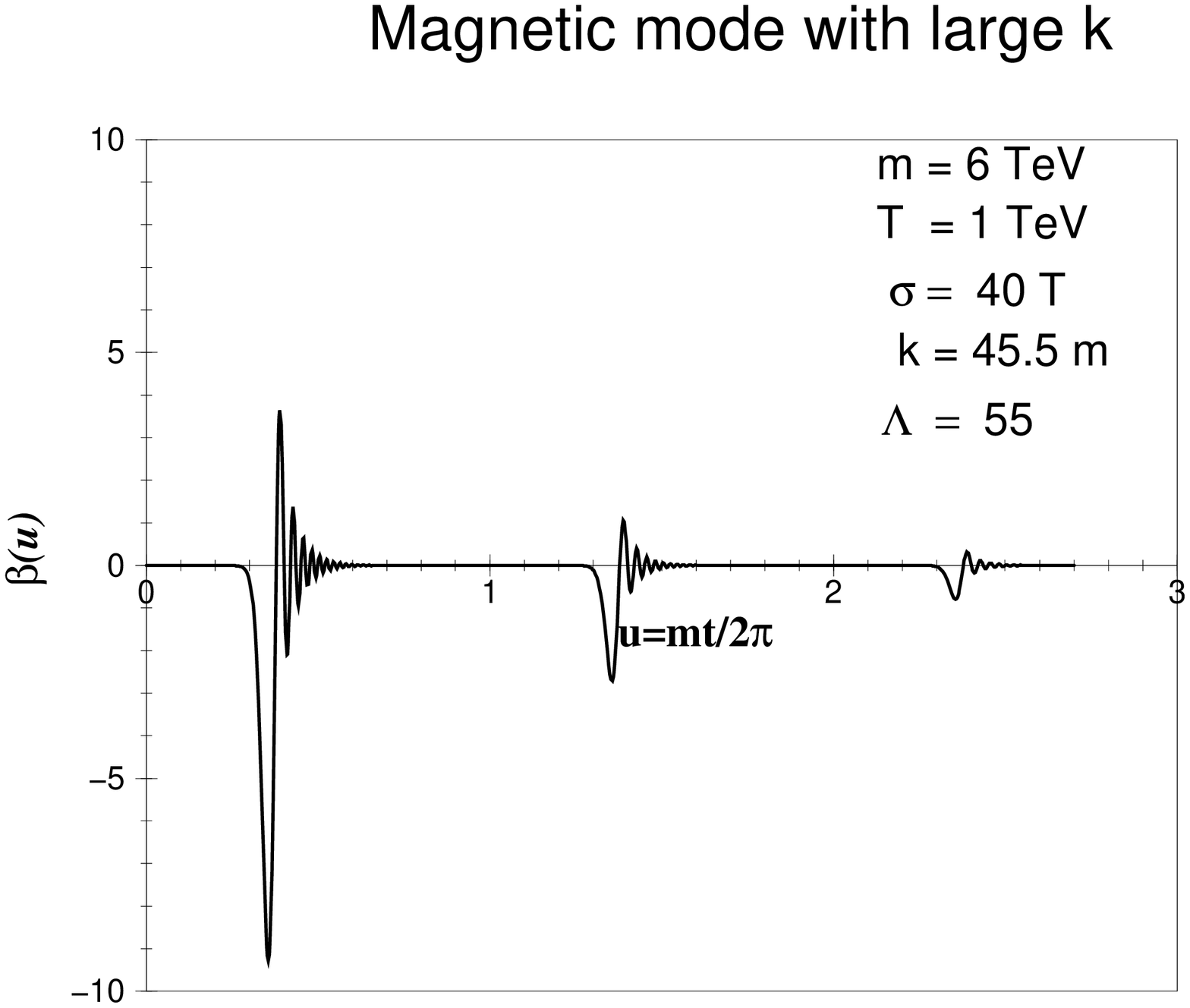,height=2.7in}\\

\

\psfig{figure=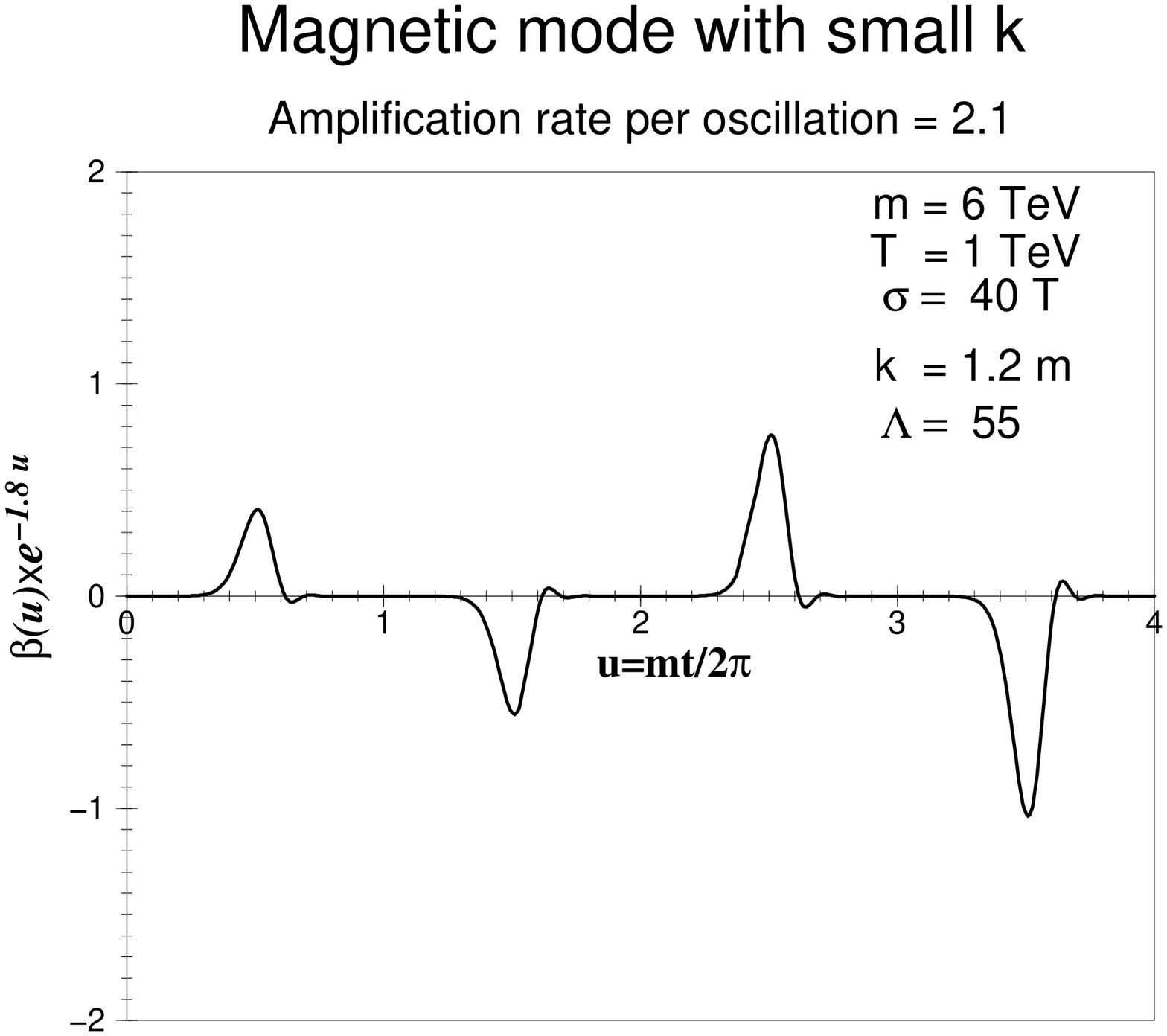,height=2.7in}\\

\

\psfig{figure=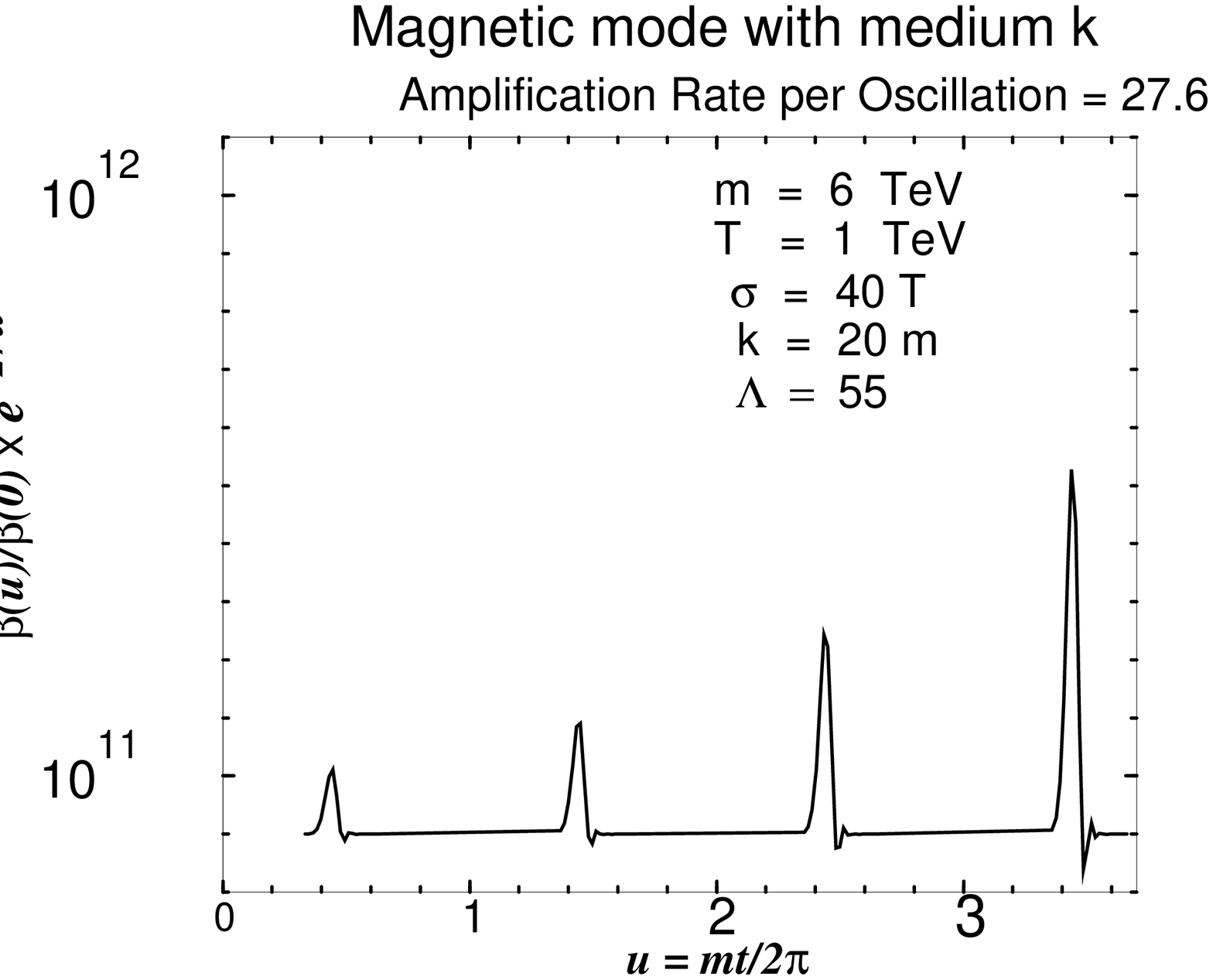,height=2.7in}

\end{center}
\caption{Amplification of EM fields. The function $\beta$, exponentially
scaled,  is shown for the parameters $\Lambda=55$, $m=6 TeV$ and
$\sigma=40 T$ at $T=1 TeV$, for three different representative wave
numbers.
\label{fig:amp} }
\end{figure}

This mechanism is a
very efficient amplifier of EM fields. For example, to obtain an amplification
of $10^{12}$ for $\Lambda=55$, $k/m=20$, $m=6 TeV$ and $\sigma=40 T$, for
oscillations occurring at a temperature of $1 TeV$, we need just one cycle!
Other examples: for $\Lambda=18$, $k/m=2$, $m=6 TeV$ and $\sigma=70 T$
and oscillations occurring at $T=120 GeV$, in one cycle magnetic fields
are amplified by a factor $10^{4}$; for $\Lambda=50$, $k/m=10$, $m=6 TeV$
and $\sigma=40 T$  at $T=120 GeV$, the amplification factor is $10^{17}$.

For the range of parameters in which fields are amplified, the amount of
amplification per cycle for each of the two modes $\Gamma^{\pm}$ is
very well approximated by the same  constant $\Gamma(k/m,\Lambda,\sigma)$.
A good approximate estimate for the average amplification after $N$
cycles is therefore 
${{\cal A}^\pm}(k,\eta)={{\cal N}^\pm}_{k} e^{N \Gamma}$, where
${{\cal N}^\pm}_{k}$ represents the transient influence of the initial
conditions of EM and scalar fields.  

In Figures \ref{fig:amppcwn} and \ref{fig:amppct} we have 
shown an example of the claimed dependence of
$\Gamma$ on wave number and temperature,

\begin{figure}
\begin{center}
\ \\

\psfig{figure=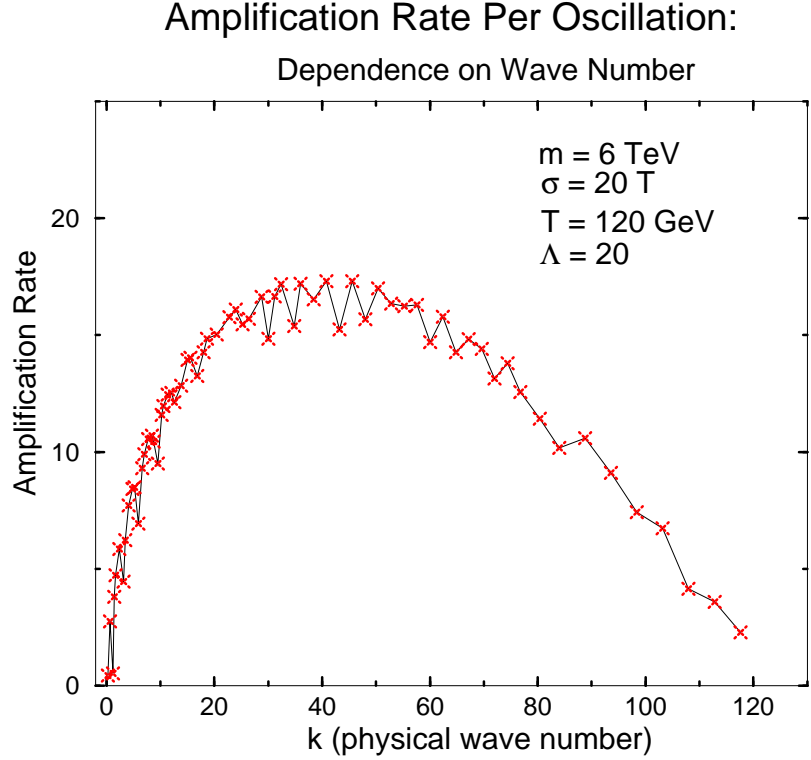,width=5in}

\end{center}
\caption{Amplification per cycle of EM fields, as a function of
wave number.
 \label{fig:amppcwn} }
\end{figure}

\begin{figure}
\begin{center}
\psfig{figure=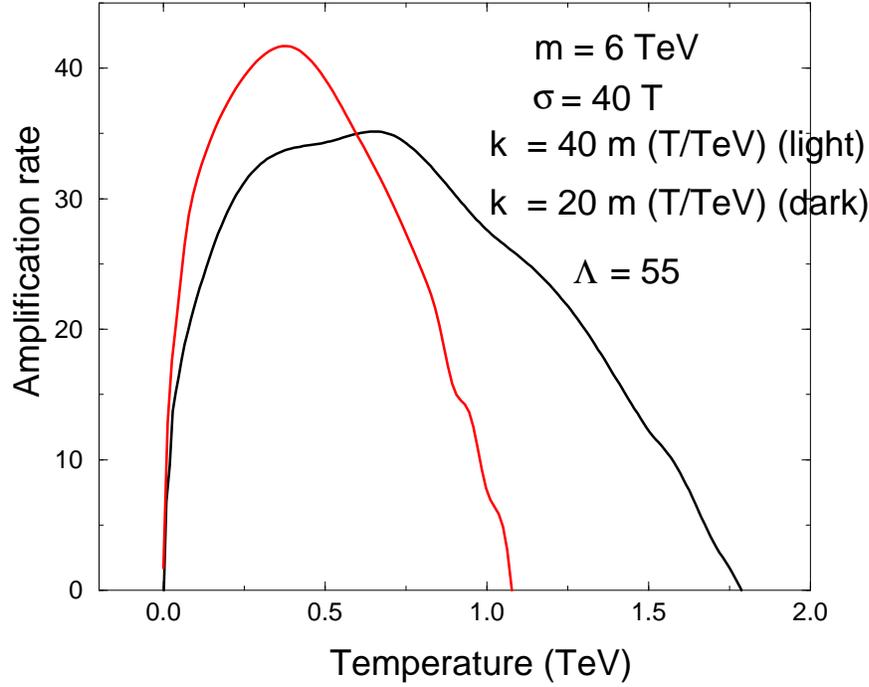,,width=5in}
\end{center}
\caption{Amplification per cycle of EM fields, as a function of
temperature.
 \label{fig:amppct} }
\end{figure}

In Fig. \ref{fig:amppcg} we represent the
amplification rate per oscillation as a function of the amplitude of the
scalar oscillations $\Lambda$. We show the 
range of variation with the conductivity $\sigma$ for 
range $\sigma = 10-70 T$. A
notable feature of this picture are the dips for certain values of
$\Lambda$ in both  graphs. We believe that these are specific values, 
for which through the coupling of higher Fourier modes of 
the periodic function ${\it P}(\eta)$ in ({\ref {perioampl}}) 
to the scalar oscillation the leading exponential factor is 
canceled. But we do not have a clear understanding of
this phenomenon.

\begin{figure}
\begin{center}
\psfig{figure=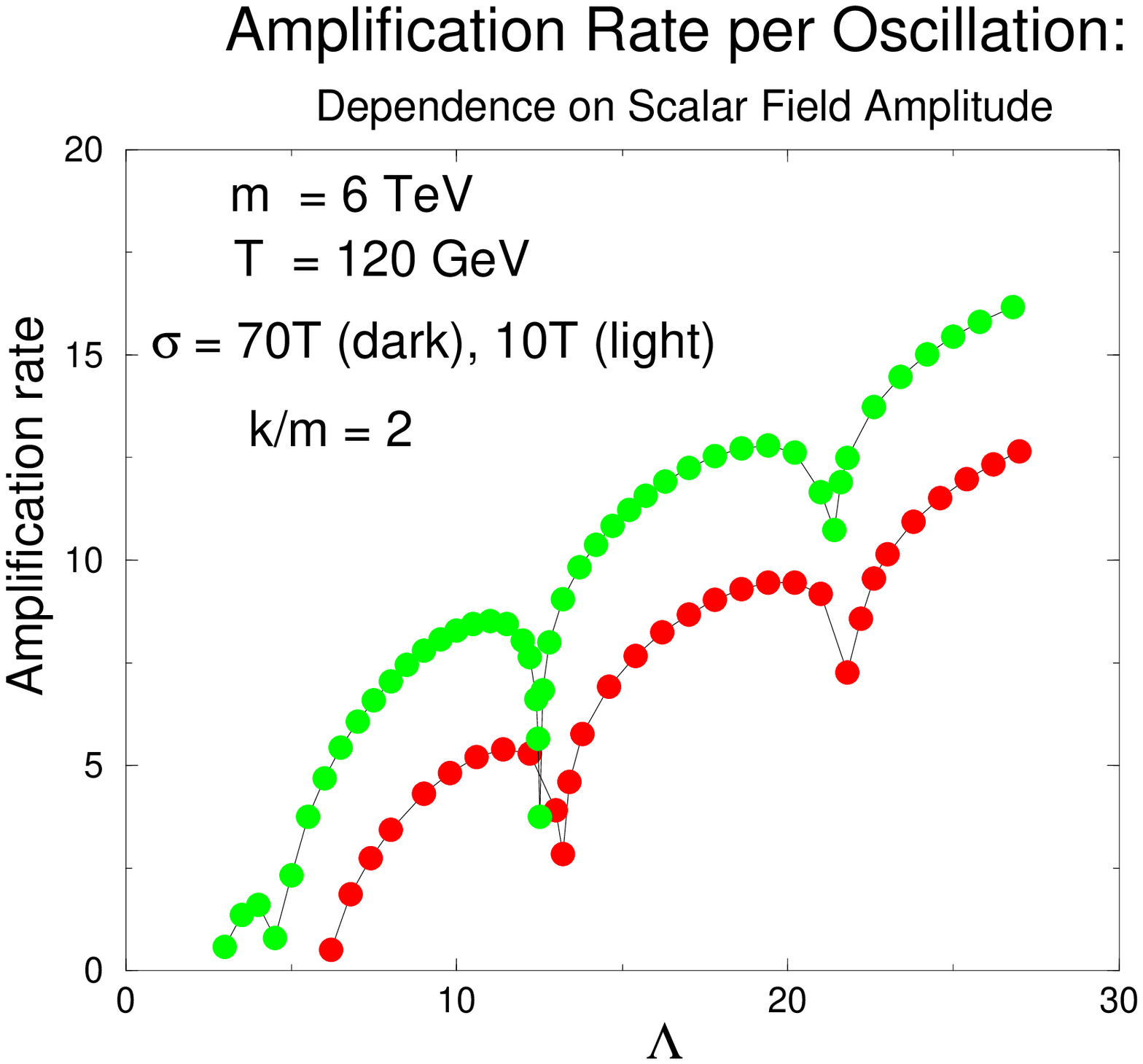,width=5in}
\end{center}
\caption{Amplification per cycle of EM fields, as a function of 
scalar field amplitude.
 \label{fig:amppcg} }
\end{figure}

\section{CONCLUSIONS }

A pseudoscalar field coupled to the hypercharge topological
density can exponentially amplify HEM fields
and develop a net Chern-Simons number in the symmetric phase of the EW
plasma  while it rolls or oscillates around the
minimum of its potential. This mechanism
could drastically change the electroweak scenario for baryogenesis and
perhaps fix its two main dissabilities: the amount of asymmetry
generated by electroweak processes and the character of the phase
transition.

We have studied this mechanism for (pseudo)scalar masses in the TeV range,
that could naturally appear if the scalar field
is associated to supersymmetry breaking. In    
that case the coherent oscillations have to occur just before or during  
the phase transition in order to avoid the fast diffusion in the plasma
of amplified magnetic modes once the scalar coherent motion terminates. In such a case the amplification
spectrum is sharply peaked around the wave number $k_{\it max} \sim
\frac{1}{2} \Lambda m$, total amplification is exponential in the number
of cycles, and  large amplification of magnetic modes, even $10^{12}$ or
larger, can happen just after a few scalar oscillations depending on the 
particular values of the parameters of the model.

If the scalar field rolls instead of oscillating, the mechanism would be
relevant for EW baryogenesis even if it takes place
 at higher temperatures
 before the phase transition. In this case, modes with wave number  much
smaller than the temperature of the plasma $k \ll T$ are maximally
amplified. Once the scalar rolling terminates, the amplified magnetic modes
remain frozen in the plasma and do not diffuse. 

In a previous paper we have concluded that our
 mechanism would be able to
generate enough asymmetry to explain the baryon number density to entropy
ratio observed in the universe. 

\acknowledgments 
This work is supported in part by the  Israel
Science Foundation administered by the Israel Academy of Sciences and
Humanities. D.O. is supported in part by the Ministry of Education and
Science of Spain. We thank Eduardo Guendelman for discussions.

\end{document}